\documentclass[letter,prd,aps,floatfix,superscriptaddress,nofootinbib]{revtex4}
\usepackage{lipsum}
\usepackage{amssymb,amsmath,graphicx}
\usepackage{hyperref}
\usepackage[usenames,dvipsnames]{color}
\usepackage{array}
\usepackage{footnote}

\makeatletter
\renewcommand*{\p@section}{\S\,}
\renewcommand*{\p@subsection}{\S\,}

\makeatother

\def\jnl@style{\it}
\def\aaref@jnl#1{{\jnl@style#1}}

\def\aaref@jnl#1{{\jnl@style#1}}

\def\aj{\aaref@jnl{AJ}}                   
\def\araa{\aaref@jnl{ARA\&A}}             
\def\apj{\aaref@jnl{ApJ}}                 
\def\apjl{\aaref@jnl{ApJ}}                
\def\apjs{\aaref@jnl{ApJS}}               
\def\ao{\aaref@jnl{Appl.~Opt.}}           
\def\apss{\aaref@jnl{Ap\&SS}}             
\def\aap{\aaref@jnl{A\&A}}                
\def\aapr{\aaref@jnl{A\&A~Rev.}}          
\def\aaps{\aaref@jnl{A\&AS}}              
\def\azh{\aaref@jnl{AZh}}                 
\def\baas{\aaref@jnl{BAAS}}               
\def\jrasc{\aaref@jnl{JRASC}}             
\def\memras{\aaref@jnl{MmRAS}}            
\def\mnras{\aaref@jnl{MNRAS}}             
\def\pra{\aaref@jnl{Phys.~Rev.~A}}        
\def\prb{\aaref@jnl{Phys.~Rev.~B}}        
\def\prc{\aaref@jnl{Phys.~Rev.~C}}        
\def\prd{\aaref@jnl{Phys.~Rev.~D}}        
\def\pre{\aaref@jnl{Phys.~Rev.~E}}        
\def\prl{\aaref@jnl{Phys.~Rev.~Lett.}}    
\def\pasp{\aaref@jnl{PASP}}               
\def\pasj{\aaref@jnl{PASJ}}               
\def\qjras{\aaref@jnl{QJRAS}}             
\def\skytel{\aaref@jnl{S\&T}}             
\def\solphys{\aaref@jnl{Sol.~Phys.}}      
\def\sovast{\aaref@jnl{Soviet~Ast.}}      
\def\ssr{\aaref@jnl{Space~Sci.~Rev.}}     
\def\zap{\aaref@jnl{ZAp}}                 
\def\nat{\aaref@jnl{Nature}}              
\def\iaucirc{\aaref@jnl{IAU~Circ.}}       
\def\aplett{\aaref@jnl{Astrophys.~Lett.}} 
\def\apspr{\aaref@jnl{Astrophys.~Space~Phys.~Res.}}
\def\bain{\aaref@jnl{Bull.~Astron.~Inst.~Netherlands}} 
\def\fcp{\aaref@jnl{Fund.~Cosmic~Phys.}}  
\def\gca{\aaref@jnl{Geochim.~Cosmochim.~Acta}}   
\def\grl{\aaref@jnl{Geophys.~Res.~Lett.}} 
\def\jcp{\aaref@jnl{J.~Chem.~Phys.}}      
\def\jgr{\aaref@jnl{J.~Geophys.~Res.}}    
\def\jqsrt{\aaref@jnl{J.~Quant.~Spec.~Radiat.~Transf.}}
\def\memsai{\aaref@jnl{Mem.~Soc.~Astron.~Italiana}}
\def\nphysa{\aaref@jnl{Nucl.~Phys.~A}}   
\def\physrep{\aaref@jnl{Phys.~Rep.}}   
\def\physscr{\aaref@jnl{Phys.~Scr}}   
\def\planss{\aaref@jnl{Planet.~Space~Sci.}}   
\def\procspie{\aaref@jnl{Proc.~SPIE}}   

\begin{document}

\date{\today}
\title{Stability and Observability of Magnetic Primordial Black Hole-Neutron Star Collisions}

\author{John Estes}
\affiliation{Department of Chemistry and Physics, SUNY Old Westbury, Old Westbury, NY, United States}
\author{Michael Kavic}
\affiliation{Department of Chemistry and Physics, SUNY Old Westbury, Old Westbury, NY, United States}
\author{Steven L. Liebling}
\affiliation{Department of Physics, Long Island University, Brookville, NY, United States}
\author{Matthew Lippert}
\affiliation{Department of Chemistry and Physics, SUNY Old Westbury, Old Westbury, NY, United States}
\author{John H. Simonetti}
\affiliation{Department of Physics, Virginia Tech, Blacksburg, VA, United States}



\begin{abstract}
The collision of a primordial black hole with a neutron star results in the black hole eventually consuming the entire neutron star.  However, if the black hole is magnetically charged, and therefore stable against decay by Hawking radiation, the consequences can be quite different.  Upon colliding with a neutron star, a magnetic black hole very rapidly comes to a stop.  For large enough magnetic charge, we show that this collision can be detected as a sudden change in the rotation period of the neutron star, a glitch or anti-glitch.We argue that the magnetic primordial black hole, which then settles to the core of the neutron star, does not necessarily devour the entire neutron star; the system can instead reach a long-lived, quasi-stable equilibrium.  Because the black hole is microscopic compared to the neutron star, most stellar properties remain unchanged compared to before the collision. However, the neutron star will heat up and its surface magnetic field could potentially change, both effects potentially observable.

\end{abstract}

\maketitle

\tableofcontents

\section{Introduction}\label{introduction}

The first direct detection of gravitational waves by LIGO and Virgo was followed just a couple years later by the first concurrent detection of a source in both gravitational and electromagnetic bands, heralding a golden age of multimessenger astronomy. Just recently LIGO and Virgo detected the first black hole~(BH)~merging with a neutron star~(NS)~\cite{LIGOScientific:2021qlt}.  In fact, LIGO and Virgo detected two such mergers, both in a regime in which electromagnetic counterparts are not expected (and none were observed). For such mergers, when the black hole is much more massive than a neutron star, the hole tends to devour the neutron star whole without much disrupted material to source an observable electromagnetic signal. 

The outcome of a BH-NS merger in the opposite regime, in which the neutron star is much more massive than the black hole, is less straightforward. Although black holes resulting from stellar collapse are invariably at least as large as a NS, primordial black holes~(PBHs) can be created by large density fluctuations in the early universe at any mass, in principle down to the Planck scale (see \cite{Sasaki:2018dmp, Green:2020jor, Villanueva-Domingo:2021spv} for recent reviews). While the idea of PBHs is not new, the suggestion that large black holes recently observed by LIGO/Virgo may have primordial origin \cite{Bird:2016dcv} has spurred renewed interest.

PBHs are a possible dark matter component, subject to a number of phenomenological bounds. As a result of Hawking radiation, very light PBHs, with $M_\mathrm{PBH} < 5\times 10^{11} \, \rm kg$, would have completely evaporated by now,\footnote{The end stage of the BH evaporation process is poorly understood, and certain proposals for resolving the BH information problem suggest that, rather than completely evaporating, the BH could evolve into some sort of very light remnant.  For a review, see \cite{Chen:2014jwq}.} and somewhat more massive PBHs with $M_\mathrm{PBH} < 10^{14} \, \rm kg$ would still emit sufficient radiation that their abundance is quite limited. The abundance of PBHs in the planetary to stellar mass range, $10^{19}\, {\rm kg} < M_\mathrm{PBH} < 10^{33} \, {\rm kg}$, is modestly constrained by the frequency of stellar microlensing events, though for stellar mass PBHs, $M_\mathrm{PBH} > 5\times 10^{30} \, \rm kg$, dynamical effects, limits on radiation from accreted gas, and gravitational wave signals impose severe limits.  

The sub-planetary mass range, $10^{14} \, {\rm kg} < M_\mathrm{PBH} < 10 ^{19} \,  {\rm kg}$, however, is still quite unconstrained. Recent work has argued that NANOgrav may have observed stochastic gravitational wave signals from PBH formation in the early universe, suggesting that PBHs comprise a significant fraction of dark matter~\cite{DeLuca:2020agl,Vaskonen:2020lbd}.

If a small PBH collides with a NS, it falls to the center of the NS due to dynamical friction and begins accreting dense nuclear matter, eventually devouring the entire NS.  Numerical simulations~\cite{East:2019dxt, Richards:2021upu,Schnauck:2021hlm} of this process demonstrated that, at least while the BH is still small compared to the NS, the accretion rate is proportional to the BH horizon area, indicating a Bondi-like accretion process.  These simulations assumed the black holes were large enough, however, so the effects of Hawking radiation could be neglected.

For small black holes, however, Hawking radiation can be significant.  The Hawking temperature of a Schwarzschild black hole is inversely proportional to its mass, implying a negative specific heat and an instability to evaporation.  However, for a BH inside a NS, as long as the initial accretion rate is larger than the initial Hawking luminosity, the BH will grow, cool, and eventually consume the NS. This will be the case for any PBH-NS merger unless the PBH was coincidentally just about to completely evaporate when it collided with the NS.  

The situation, however, is quite different if the BH is charged.  Near-extremal, Reisner-Nordstrom black holes have positive specific heat and cool due via Hawking radiation into zero-temperature, extremal BHs.  While electrically charged BHs can readily discharge via nucleation of electron-positron pairs (or other light, charged particles), a magnetically charged BH could be very long-lived, because the large mass of magnetic monopoles exponentially suppresses the rate of monopole-anti-monopole pair production \cite{Maldacena:2020skw}.

A magnetized PBH~(mPBH) could result from the accretion, in the early universe, of monopoles by a PBH. For example, a PBH in a strong magnetic field could produce a pair of magnetic monopoles either through pair production~\cite{Kobayashi:2021des} by the magnetic field itself or through the Kibble-Zurek mechanism as the black hole evaporates~\cite{Das:2021wei}.  The PBH could swallow one of the magnetic monopoles becoming a mPBH, while the other is accelerated away by the strong magnetic field. The mPBH then continues to emit Hawking radiation, shedding its excess mass, and by the present epoch, a small mPBH  will have evolved very near to extremality.

Once created, a mPBH of charge $Q$ has many interesting and unusual properties compared to its electric counterpart, in addition to its stability.\footnote{We work in units where the charge $Q$ is dimensionless.} For a mPBH with $Q \lesssim 10^{32}$, the large magnetic field near the horizon causes the SU(2) gauge fields to condense, while the Higgs vacuum expectation value vanishes~\cite{Maldacena:2020skw}.  This near-horizon region is bounded by an electroweak corona with varying SU(2) gauge and Higgs fields.  Outside the corona, the SU(2) and Higgs fields take their standard vacuum expectation values.  The scale of the corona is set by the charge $Q$ and the geometric mean of the Higgs and W-boson scales.  For example, a value of $Q \sim 10^{20}$ corresponds to a corona radius of order $10^{-8} \ \rm m$.  Inside the corona, the condensates break translation symmetry in the space transverse to the magnetic field, corresponding to the formation of vortices.  Within this region, the standard model fermions are confined to the vortices and behave as massless two-dimensional particles, significantly enhancing the Hawking radiation rate.  This results in the whole region inside the corona heating up to the temperature of the black hole~\cite{Maldacena:2020skw}. 

Some recent works have discussed observable astrophysical signatures of magnetically charged BHs \cite{Bai:2020spd, Ghosh:2020tdu}.  In \cite{Bai:2019zcd}, it was suggested that mPBHs with $M_\mathrm{PBH} < 10^{6} \, \rm kg$ could be a dominant component of dark matter.  

In this paper, we investigate the collision of a small, extremal (or near-extremal) mPBH with a NS.  We first discuss the possibility that such collisions could be observed as pulsar glitches and anti-glitches.  We find that, with the current sensitivity of radio observations, only collisions with large charge $Q > 10^{23}$ would be observable. However, since we subsequently determine that, for such large mPBHs, the NS would rapidly collapse, such collisions could not explain glitches in pulsars that are observed to continue pulsing.  However, third generation gravitational wave detectors might have sufficient sensitivity to allow observations of impacts of smaller charge BHs which could lead to stable bound states.

After the collision, the mPBH will quickly settle down to the core and begin accreting the dense NS material.  Unlike the case of a Schwarzschild BH in a NS, which eventually devours the entire star, we argue that a captured mPBH can evolve to a long-lived, quasi-stationary state. The mPBH heats up as it accretes infalling matter, then re-emits the energy via Hawking radiation.  When the outgoing radiation flux balances the infalling mass flux, the mPBH stops growing and reaches stable equilibrium.  Hawking radiation provides the central pressure to support the NS.  If the magnetic charge of the BH is not too large, $Q < 10^{17}$, the accretion rate can increase until the Hawking pressure is sufficient to support the NS, and this mPBH-NS state will then persist until an appreciable fraction of the NS mass has been accreted and re-radiated by the black hole, which implies a lifetime for small magnetic charge $Q < 10^{10}$ that is longer than the age of the universe.

In Section~\ref{sec:collision} below, we analyze the initial collision between a mPBH and a NS, in particular how the rotation of the NS is affected.  We then discuss the fate of the mPBH when it reaches the NS core and the possibility of forming a long-lived, quasi-stationary state in Section~\ref{sec:PBHns}. We analyze the process by which the mPBH acretes NS material and then re-radiates the energy as Hawking radiation in Section~\ref{sec:BH_equil}.  Then in Section~\ref{sec:ode}, we investigate the evolution and stability of the NS, and in Section~\ref{sec:scenarios} enumerate three qualitative scenarios based on the mPBH charge. In Section~\ref{sec:observables} we discuss possible ways to observationally distinguish neutron stars with and without an embedded mPBH, followed by a discussion of open questions in Section~\ref{sec:discussion}.

\section{Collision of a magnetized PBH with a Pulsar} \label{sec:collision}

An uncharged PBH colliding with, or even passing sufficiently near, a NS will lose kinetic energy through dynamical friction, accretion of NS matter, and emission of gravitational waves \cite{Capela:2013yf, Defillon:2014wla, Genolini:2020ejw}.  The PBH will get captured eventually, but only after repeated passes through the NS, with a light BH of mass $m_\mathrm{PBH} < 10^{19} \, \rm kg$ 
 taking more than $10^4$~years \cite{Genolini:2020ejw}.

For a magnetically charged PBH, on the other hand, electromagnetic interactions with the NS are dominant, and even a relativistic mPBH will be rapidly and efficiently captured \cite{Bai:2020spd}.  

If a magnetized PBH collides with a NS which is observed as a pulsar, one might be able to detect the resulting change in the pulse period due to the rapid change in the rotation rate of the NS. Such sudden changes in the pulsar's period are observed as glitches (for a discussion of observed glitches see, e.g., \cite{Espinoza2011}). Glitches refer to sudden decreases in the pulse period, but anti-glitches, or sudden increases in pulse period, have also been observed, e.g., \cite{Archibald2013}. Huang and Geng \cite{Huang2014} consider a collision of a solid body with the neutron star as the possible cause of an anti-glitch. We consider a similar scenario here, involving impact by a mPBH.

\begin{figure}
\includegraphics[width=0.72\textwidth]{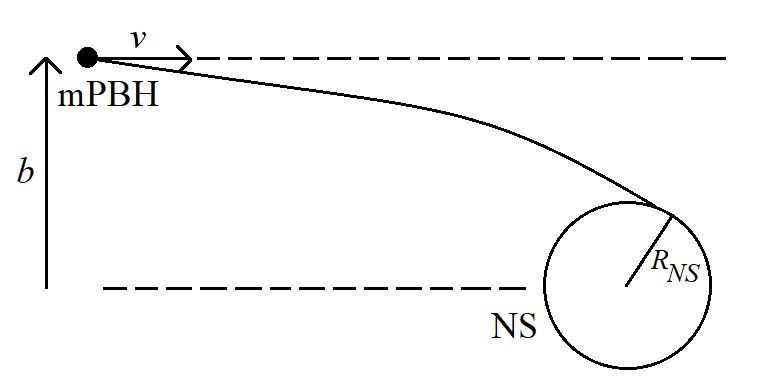}
\caption{The scenario for a mPBH colliding with a NS. The impact parameter for the mPBH is $b$, but the mPBH collides at a closest approach of distance $\lesssim R_\mathrm{NS}$, the radius of the NS. The impact parameter $b\gg R_\mathrm{NS}$. Thus the mPBH speed at the collision is much greater than its initial speed $v$ and is comparable to the escape velocity for the NS.
}\label{fig:PBHvsns}
\end{figure}

Consider an extremal mPBH of charge $Q$ moving at virial speed in the Galaxy, i.e., $v\sim10^2$~km/s $\sim10^{5}$~m/s. The mass of the mPBH is 
\begin{equation}
    M_\mathrm{BH} = \frac{\hbar c\sqrt{\pi\epsilon_0}}{e\sqrt{G_N}} Q = (1.27 \times 10^{-7}\, \rm{kg} )\, Q.
\end{equation}
While it is initially very far from the NS, it eventually collides with the NS because its impact parameter $b$ is smaller than the gravitational cross section for collision with the NS. In other words, while it would have missed the NS by distance $b$ if traveling along a straight path, the gravitational attraction of the NS causes the point of closest passage of the mPBH to the NS to be less than the radius of the NS, $R_\mathrm{NS}$. See Fig.~\ref{fig:PBHvsns}. Take the mass of the NS to be $M_\mathrm{NS}$. 

During the motion of the mPBH toward the NS, the angular momentum and mechanical energy  of the mPBH-NS system are conserved. Thus the angular momentum of the NS will be changed by an amount of magnitude $M_\mathrm{BH}vb$ when the NS absorbs the mPBH.  To determine $M_\mathrm{BH}vb$ note that
\begin{equation}
    M_\mathrm{BH}vb\sim m v_c R_\mathrm{NS}
\end{equation}
where $v_c$ is the speed of the mPBH when it collides with the NS. In obtaining this result we have assumed that the $M_\mathrm{BH}vb$ is much less than the angular momentum of the rotating NS. In other words, the change in the angular momentum of the NS will be much less than its initial angular momentum. This assumption will be proven correct below.

The escape velocity for a NS is 
\begin{equation}
    v_\mathrm{esc}=\sqrt{\frac{2GM_\mathrm{NS}}{R_\mathrm{NS}}}\sim10^8 \ {\rm m/s},
\end{equation}
for $M_\mathrm{NS}\sim M_\odot \sim 10^{30}$~kg, and $R_\mathrm{NS}\sim 10^4$~m. Since the mPBH starts very far from the NS with speed $v\ll v_\mathrm{esc}$, it collides with the NS with speed very close to $v_\mathrm{esc}$. Thus the magnitude of the angular momentum imparted to the NS is
\begin{equation}
    M_\mathrm{BH}vb\sim M_\mathrm{BH}v_\mathrm{esc}R_\mathrm{NS}.
\end{equation}
The fractional change in the angular momentum of the NS is therefore
\begin{equation}
    \frac{M_\mathrm{BH}v_\mathrm{esc}R_\mathrm{NS}} {I_\mathrm{NS}\omega_\mathrm{NS}} \sim \frac{M_\mathrm{BH}v_\mathrm{esc}P_\mathrm{rot}} {2\pi M_\mathrm{NS}R_\mathrm{NS}}
    \sim \frac{Q}{10^{24}} 10^{-10}
\end{equation}
where $I_\mathrm{NS}$ and $\omega_\mathrm{NS}$ are the moment of inertia and angular rotation rate of the NS, respectively, and
we have assumed the rotation period of the NS is about 1 second. This result for the fractional change justifies our assumption that the change in the angular momentum of the NS will be very small. Since the fractional change in the moment of inertia of the NS, due to the embedding of the mPBH, is roughly
\begin{equation}
    \frac{M_\mathrm{BH}R_\mathrm{NS}^2}{M_\mathrm{NS}R_\mathrm{NS}^2}\sim \frac{Q}{10^{24}}10^{-13},
\end{equation}
and is much smaller than the fractional change in angular momentum, then the fractional change in the rotation period of the NS will also be
\begin{equation}
    \frac{\Delta P_\mathrm{rot}}{P_\mathrm{rot}}
    \sim \frac{Q}{10^{24}} 10^{-10}.
\end{equation}

Current observations show typical glitches have $\Delta P_\mathrm{rot}/P_\mathrm{rot}\sim 10^{-11} - 10^{-5}$ \cite{2017A&A...608A.131F}. If at least some of these glitches are the result of mPBH-NS collisions, then they imply 
$Q\sim 10^{23} - 10^{29}$. As shown in Sec.~\ref{sec:PBHns} below, if such an mPBH is absorbed by a NS, the result would be unstable. The pulsar would soon be unobservable. Since pulsars with these observed glitches continue to pulse, we can say such events are not due to collisions with mPBHs of such large values of $Q$. 

However, glitches and anti-glitches also generate transient gravitational wave signals which may be observable by future third generation detectors \cite{Yim:2022qcn, Lopez:2022yph}.  With sufficient sensitivity in band, future gravitational wave detectors could make it possible to observe small glitches currently unobservable via pulsar emission, in which case this collision scenario may be viable.

Furthermore, one might  hope for such observations to differentiate among the various explanations for glitches, because, for the case considered here, a glitch would be immediately preceded by a (very unequal mass) binary merger. However, such a merger would be at very high frequency (a few kilohertz), a band in which detectors tend to be less sensitive~\cite{Genolini:2020ejw,Zou:2022wtp}.
The interaction of the BH's magnetic field with that of the NS also offers the potential of
an electromagnetic precursor signal, which may differentiate the event from an asteroid collision. Precursor electromagnetic signals from BH-NS mergers have been
predicted although with a couple caveats: (1)~such studies considered stellar sized BHs, not PBHs and (2)~they assume an uncharged BH~\cite{DOrazio:2021puy,Carrasco:2021jja,Chen:2021sya,East:2021spd,Pan:2019ulx,Paschalidis:2013jsa,Lai:2012qe}.

\section{MPBH-NS bound states}
\label{sec:PBHns}

\subsection{Black holes in neutron stars}
\label{sec:BH_equil}

Once a mPBH is stopped and settles down to the center of the NS, it will begin to accrete material from the NS core.  For an uncharged PBH, this accretion process has been estimated analytically \cite{Giddings:2008gr, Fuller:2017uyd} and simulated numerically \cite{East:2019dxt, Richards:2021upu} and, for a small PBH (i.e. $M_\mathrm{BH} \ll M_\mathrm{NS}$), the Bondi-like accretion rate is proportional to the horizon area $A_\mathrm{BH}$ of the PBH,\footnote{East and Lehner \cite{East:2019dxt} simulate several NS equations of state, and the average accretion rate is approximate $\dot M_\mathrm{BH} \approx (1.5 \times 10^{15} \ {\rm \frac{g}{cm^3}}) \ A_\mathrm{PBH}$ (see Fig 2 in \cite{East:2019dxt}).}
\begin{equation}
\label{eq:ELaccretion}
    \dot M_\mathrm{BH} =  \left( 4.5 \times 10^{26} \ {\rm \frac{kg}{m^2 s}} \right) \ A_\mathrm{BH}  \ .
\end{equation}
As the PBH grows, the accretion accelerates until the entire NS has been consumed, leaving a stellar mass BH in its place.

The result of capturing a mPBH is quite different.  An extremal mPBH with magnetic charge $Q$ is metastable\footnote{An extremal mPBH can decay by nucleation of monopole-anti-monopole pairs, a process which is suppressed by the high mass of a monopole.} and has zero temperature.  Once it begins accreting, the temperature will rise and the mPBH will begin emitting Hawking radiation.  A non-extremal mPBH with mass $M_\mathrm{BH}$ has a temperature is given by
\begin{align}
\label{HawkingTemperature_full}
    T = \frac{\hbar c^3}{2 \pi k_B G} \frac{\sqrt{M_\mathrm{BH}^2 - M_e^2}}{\left(M_\mathrm{BH}+\sqrt{M_\mathrm{BH}^2-M_e^2}\right)^2} \ ,
\end{align}
where $M_e = \frac{\hbar c\sqrt{\pi\epsilon_0}}{e\sqrt{G_N}} Q = (1.27 \times 10^{-7}\, \rm{kg} )\, Q$ is the mass of the original extremal mPBH. We define a dimensionless non-extremality parameter $x \equiv \Delta M/M_e = (M_\mathrm{BH}-M_e)/M_e$, which allows the temperature to be written as
\begin{align}
\label{HawkingTemperature}
    T = \left(3.9 \times 10^{30} \ \rm K \right) \frac{F(x)}{Q} \ ,
\end{align}
where the dimensionless function $F(x)$ is
\begin{equation}
    F(x) = \frac{\sqrt{x^2+2x}}{(1+x + \sqrt{x^2+2x})^2}.
\end{equation}
For a near-extremal mPBH, the temperature grows with mass as $\sqrt{M_\mathrm{BH}-M_e}$, or as $x^{1/2}$, yielding a positive specific heat.  However, for $M_\mathrm{BH} \gg M_e$, the black hole resembles a Schwarzschild BH, with a temperature decreasing as $M_\mathrm{BH}^{-1}$ and, consequently, a negative specific heat.

For a near-extremal mPBH with $Q < 3 \times 10^{32} = Q_\mathrm{ew}$, the magnetic field just outside the horizon is sufficiently large that the electroweak gauge bosons condense and the electroweak symmetry is restored.  As a result, in this region all standard model fermions are massless, and the low-energy modes are constrained by the strong magnetic field to radial motion along the field lines. The number of such two-dimensional massless modes scales as $Q$ and the Hawking radiation is essentially 1+1 dimensional.

The local magnetic field decreases with distance from the mPBH, until the field strength is equal to the square of the Higgs mass, at which point electroweak symmetry begins breaking across a transition region called the ``electroweak corona".  Moving radially through the corona, the Higgs vacuum expectation value increases from zero to its usual value, and outside is the ordinary electroweak vacuum.  Although this transition is not especially sharp and the corona is not spherically symmetric, the approximate radial location of the corona is $r_\mathrm{C} = (1.7 \times 10^{-18}\ {\rm m} ) \sqrt{Q} \ $.

This mass barrier acts as a potential, trapping the standard model fermions within the corona.  When the Hawking radiation reaches the electroweak corona, the standard model fermions are reflected back.  The whole region inside the corona heats up to the Hawking temperature set by the blackhole horizon, and the corona radiates like a thermal blackbody.

\begin{figure}
    \centering
    \includegraphics[width=0.72\textwidth]{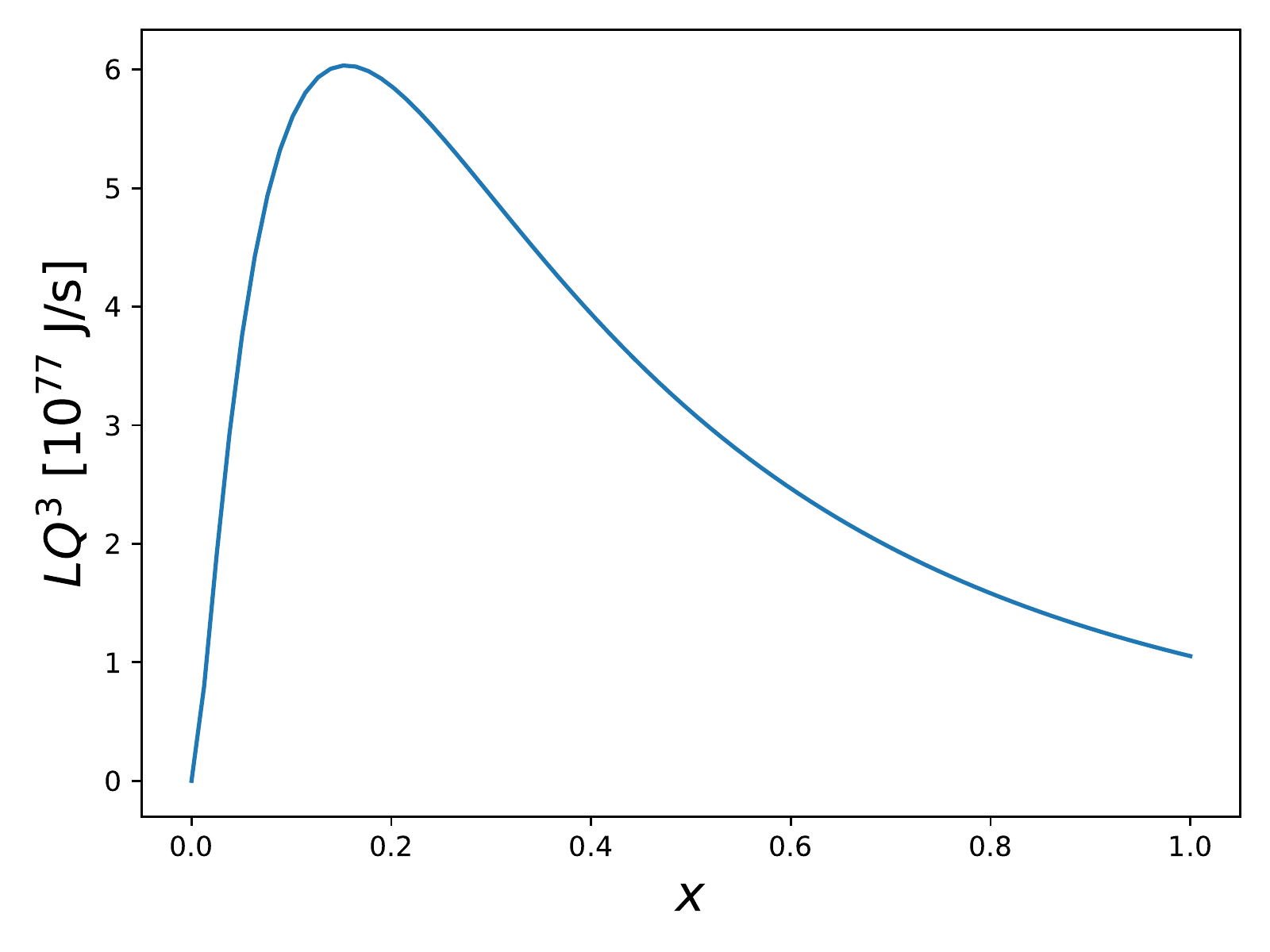}
    \caption{The Hawking flux $L$ for a mPBH, scaled by $Q^3$, as a function of the non-extremalty parameter $x$, given by  Eq.~\ref{eq:HawkingPower}. The flux reaches a maximum at $x_c = 0.15$.  For a near-extremal mPBH, $x<x_c$, the positive slope indicates thermodynamic stability.  Farther from extremality, $x >x_c$, the black hole becomes Schwarzschild-like, with a negative specific heat.}
    \label{fig:Hawkingflux}
\end{figure}

The Stephan-Boltzmann law relates the luminosity of the EW corona to the temperature
\begin{equation}
    L = A_\mathrm{C}\sigma T^4
\end{equation}
where $A_\mathrm{C}$ is the area of the corona and the Stephan-Boltzmann constant $\sigma = \frac{2\pi^5 k_B^4}{15 c^2 h} = \frac{\pi^2 k_B^4}{60 c^2 \hbar^3}$.  Inserting the expression for the Hawking temperature \eqref{HawkingTemperature} in terms of the PBH mass gives:
\begin{eqnarray}
\label{eq:HawkingPower}
  L_\mathrm{BH}  &=& \left( \frac{c^{10} \hbar }{960 \pi^2 G^4} \right)  A_\mathrm{C} T^4  \\
    &=& \left(1.2 \times 10^{115}\ {\rm \frac{J}{m^2s}}\right) \ A_\mathrm{C} \, \frac{ F(x)^4}{Q^4} = \left(4.4 \times 10^{80}\ {\rm \frac{J}{m^2s}}\right) \ \frac{F(x)^4}{Q^3}
\end{eqnarray}
Fig. \ref{fig:Hawkingflux} shows the Hawking flux \eqref{eq:HawkingPower} plotted as a function of $x$. For a near-extremal mPBH, where $x \ll 1$, the Hawking temperature and radiation flux grows with $x$ and the mPBH has a positive heat capacity. However, the power output reaches a maximum at a critical value of $x_c = 0.155$ where the function $F(x)$ reaches a maximum: $F(x_c) = 0.192$. For $x>x_c$, the temperature and power fall as the mass increases, implying a negative heat capacity akin to the behavior of a Schwarzschild black hole.

An accreting mPBH at the center of a NS can rapidly reach an equilibrium in which energy is emitted as Hawking radiation at the same rate that energy falls in, in the form of NS matter.  For a given accretion rate $\dot E_\mathrm{acc}$, the condition for zero net energy flux is $\dot E_\mathrm{acc} = L_\mathrm{BH}$, or
\begin{equation}
\label{eq:equilibrium_condition}
    \dot E_\mathrm{acc} =  \left(4.4 \times 10^{80}\ {\rm \frac{J}{m^2s}}\right) \ \frac{F(x)^4}{Q^3}
\end{equation}
which should be understood as an implicit equation that dynamically determines the equilibrium value of the non-extremality parameter, $x_\mathrm{eq}$.

There are, in fact, two solutions to \eqref{eq:equilibrium_condition}, $x_{\mathrm{eq},1} < x_c$ and $x_{\mathrm{eq},2} > x_c$. At $x_{\mathrm{eq},1}$, $T'(x_{\mathrm{eq},1})>0$ implying a positive  heat capacity.  If some extra mass falls in to the black hole, the temperature increases and the Hawking emission increases to compensate, returning the system to equilibrium.  The larger solution $x_2$ has a negative heat capacity and is unstable to cooling by additional accretion.  When an extremal mPBH begins accreting at the center of the NS, the mass increases from $x=0$ and the system is expected to stabilize at $x_{\mathrm{eq},1}$.

There is, however, an upper bound on the Hawking radiation flux \eqref{eq:HawkingPower} for a given mPBH charge, as can be seen in Fig.~\ref{fig:Hawkingflux}, and if the accretion rate is above this maximum, there is no value of $x$ which satisfies \eqref{eq:equilibrium_condition}. The function $F(x)$ has a maximum value of $0.192$, which implies an upper bound on the accretion rate which can yield an equilibrium
\begin{equation}
\label{eq:equilibrium_bound}
    \dot E_\mathrm{acc}  <  \left(6.2 \times 10^{77}\ {\rm \frac{J}{m^2s}}\right) \ Q^{-3}.
\end{equation}
A mPBH which does not satisfy this condition will not be able to re-emit all the energy that falls in and will grow without bound until the entire NS is accreted.

\subsection{Neutron stars containing black hole}
\label{sec:ode}

The NS responds to the addition of an accreting, radiating mPBH to its core.  For a small enough BH, the NS heats up but is not drastically affected, and the system can settle into a new equilibrium solution.

The NS can be effectively modeled as a spherically symmetric, hydrodynamical system in terms of the stellar pressure $P(r)$, the energy density $\rho(r)$, and the mass aspect function $m(r)$, which gives the mass inside a radius $r$. The conditions for hydrostatic equilibrium given by the TOV equations:\footnote{For a review of the TOV equations and NS solutions, see Ref.~\cite{Noble:2003xx}.}
\begin{eqnarray}
\frac{dm}{dr} & = & 4 \pi r^2 \rho \label{eq:ode1}\\
\frac{dP}{dr} & = & -\frac{\left(\rho+P\right)\left(m+4 \pi r^3 P\right)}{r \left(r-2m\right)}. \label{eq:ode2}
\end{eqnarray}
For the nuclear matter of the NS, we adopt a polytropic equation of state, $P=K \rho_0^\Gamma$, with polytropic index $\Gamma=2$ and  constant $K = 10^{-2} \rm \frac{m^5}{kg s^2}$, which sets the maximum NS mass to be $1.4 M_\odot$.  Here $\rho_0$ is the rest mass density in the rest frame of the fluid so that the energy density becomes
\begin{equation}
    \rho = \left( \frac{P}{K}\right)^{1/\Gamma} + \frac{P}{\Gamma-1}. \label{eq:eos}
\end{equation}

To solve for ordinary NS solutions, these first-order ODEs are integrated out from the center with boundary conditions specified at $r=0$: a central pressure, $P_0$, $P'(0)=0$, $m(0)=0$, and $m'(0)=0$.  The stellar surface is identified by the radius $R_{NS}$ at which the pressure drops below some small threshold. The mass of the NS is given by $M_\mathrm{NS} = m(R_\mathrm{NS})$. For a given $M_\mathrm{NS}$, there may be multiple solutions. We will focus on the perturbatively stable solution, which is the solution with the larger radius.

\begin{figure}
    \centering
    \includegraphics[width=0.72\textwidth]{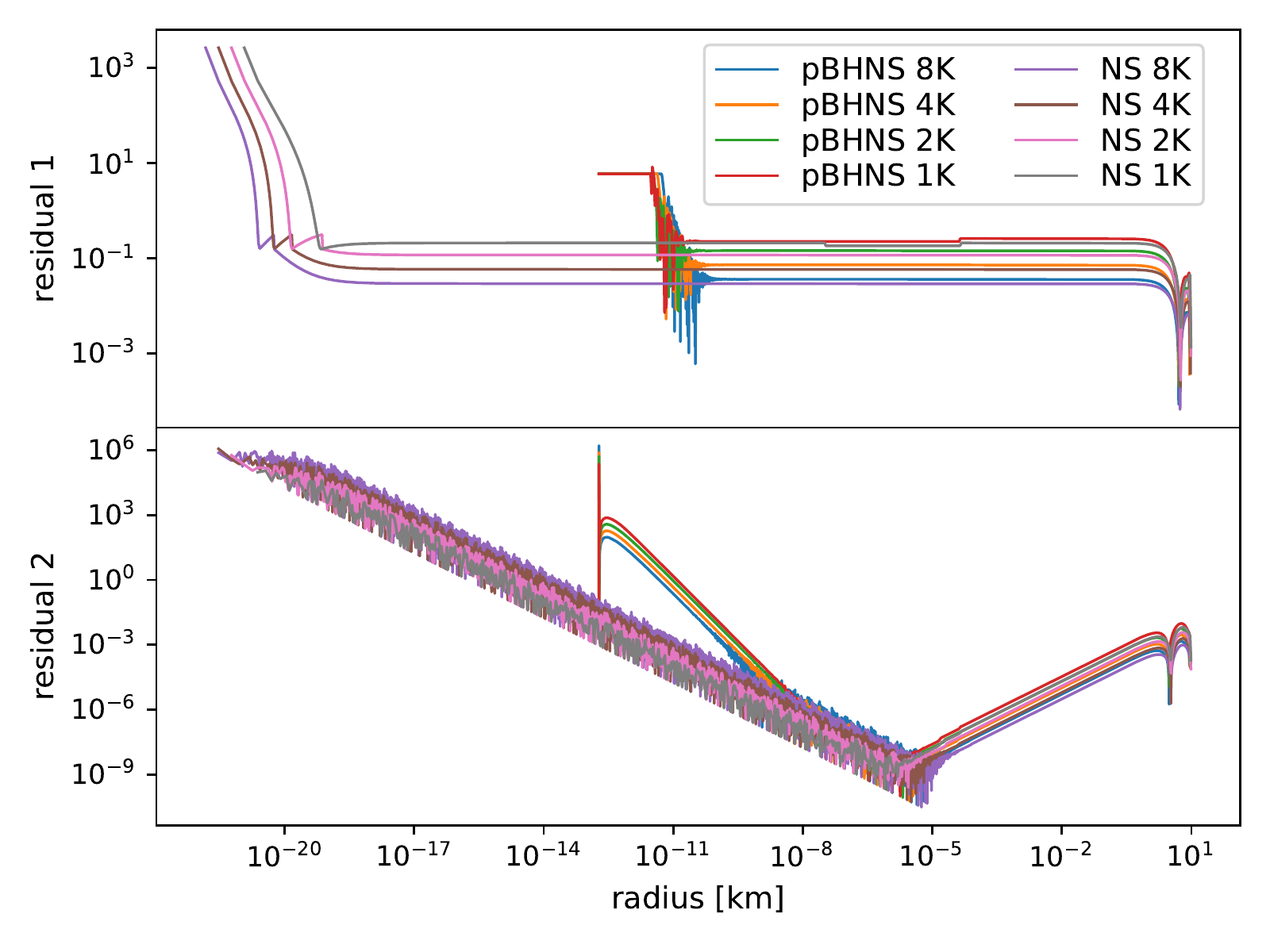}
    \caption{\textit{Convergence of numerical residuals for a particular solution.} \textbf{Top:} The residual of Eq.~\ref{eq:ode1} for a particular PBH+NS solution and similar NS solution.
    \textbf{Bottom:} The residual of Eq.~\ref{eq:ode2} for these solutions.
    The residuals for the PBH+NS generally decrease with increasing resolution, except some noisy behavior at small radius. However, this behavior is similar to that exhibited by the NS solution in which we have confidence by comparing the family of solutions with those of Ref.~\cite{Noble:2003xx}.
    }
    \label{fig:convergence}
\end{figure}

We use the \textsc{LSODA} ODE solver~\cite{etde_21352532} which automatically switches between methods for stiff and non-stiff equations. We compute numerical residuals from a particular PBH+NS solution with increasing resolution and display them in Fig.~\ref{fig:convergence}. These residuals are independent of the differencing used by the solver, and they generally converge.  There is some noisy behavior at small radius and so we also plot the residuals for a similar NS solution. We are confident in the NS solutions by comparison of the family of solutions with those of Ref.~\cite{Noble:2003xx}, and the residuals for the PBH+NS are comparable to those of the NS solution.

\begin{figure}
    \centering
    \includegraphics[width=0.72\textwidth]{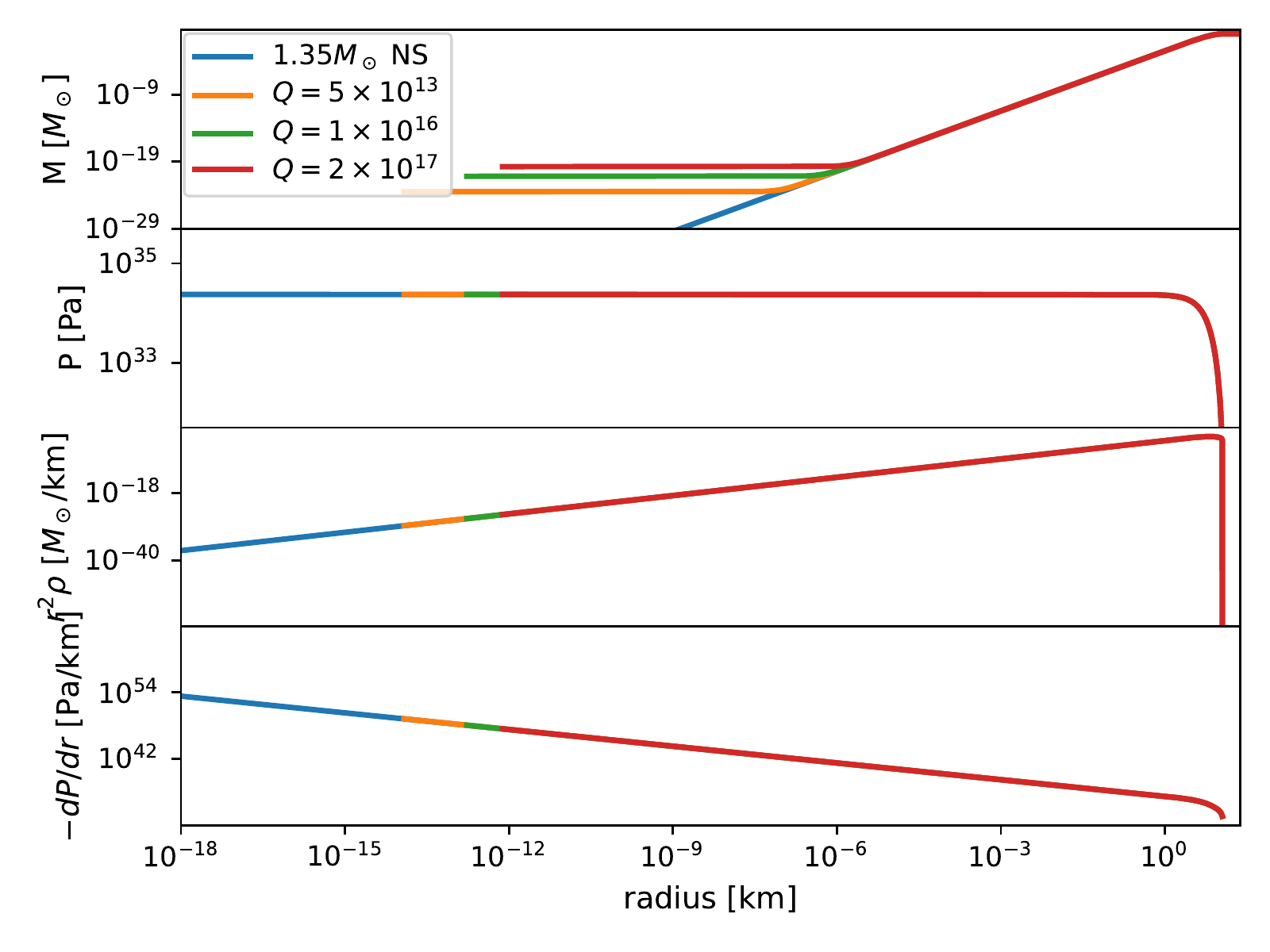}
    \caption{The fiducial $1.35 M_\odot$ NS solution and a range of mPBH-NS solutions with the same total mass but with varying mBH masses. \textbf{Top:} The mass aspect function $m(r)$. At large radius, the solutions are identical. For the mPBH-NS solutions, $m(r)$ levels off at small radius at $M_\mathrm{BH}$. 
    \textbf{Top-middle:} The pressure, $P(r)$, for the various solutions. 
    \textbf{Bottom-middle:} The energy density, $r^2\rho(r)$, for the various solutions.
    \textbf{Bottom:} The (negative) gradient of the pressure shows that the pressure is not flat at small radius.
    The curves all lie on top of each other, illustrating that, outside the corona radius, the mPBH-NS solutions have the same mass distribution as the pure NS solution.}
    \label{fig:NSsolutions}
\end{figure}

For a fiducial NS of mass $M_\mathrm{NS} = 1.35 M_\odot$, the solutions for $m(r)$ and $P(r)$ are illustrated in Fig. \ref{fig:NSsolutions}. The central pressure $P_0$ is determined by the NS mass, and along the stable branch of solutions, $P_0$ increases with $M_\mathrm{NS}$. The fiducial $1.35 M_\odot$ NS has a central pressure $P_0 = 2.37\times 10^{34}\ \rm Pa$.

We now introduce a small mPBH into the center of an existing, stable NS of mass $M_\mathrm{NS}$. In order to find NS solutions with a mPBH at the core, we modify the  boundary conditions. We begin the integration at the nonzero radius of the electroweak corona, $r_c$. At that radius, the enclosed mass is that of the extremal BH, $m(r_c)=M_\mathrm{BH}$.  The pressure at $r_c$ is now generated by the outgoing Hawking radiation, which is  
\begin{equation}
\label{eq:HawkingPresure}
   P_c \equiv P(r_c) =  \frac{L_\mathrm{BH}(x_\mathrm{eq})}{c A_c} = \left(4.0 \times 10^{106}
\ \rm Pa\right) \frac{F(x)^4}{Q^4}
\end{equation} 
where $L_\mathrm{BH}$  is the equilibrium Hawking luminosity given by \eqref{eq:HawkingPower}.  For a black hole in equilibrium, the non-extremality parameter $x$ is determined by \eqref{eq:equilibrium_condition}, and the luminosity is equal to the energy accretion rate $L_\mathrm{BH}(x_\mathrm{eq}) = \dot E_\mathrm{acc}$.

We look for solutions to \eqref{eq:ode1} and \eqref{eq:ode2} with a total mass $M_\mathrm{NS} + M_\mathrm{BH}$.  However, because we consider very small black holes, $M_\mathrm{BH} \ll M_\mathrm{NS}$, the additional mass due to the mPHB can safely be neglected. Some examples of these mPBH-NS solutions, for a range of BH masses, are shown in Fig.~\ref{fig:NSsolutions}, along with the corresponding stable NS solution of the same total mass.

The mPBH-NS solutions are very close to the pure NS solutions with a small BH placed at the center.  The outer regions of the mPBH-NS solutions, beyond the corona radius, are nearly identical to the NS solution.  Near the center of the NS, the mass aspect function is dominated by the mass of the black hole at the center.  However, as seen in the insert in Fig.~\ref{fig:NSsolutions} (Top), the NS nuclear matter extends at roughly constant density down to the corona.  Similarly, the pressures for the mPBH-NS solutions are, to good accuracy, identical to that of the pure NS, as can be seen in Fig. \ref{fig:NSsolutions} (Bottom).  Because, the mPBH-NS solutions have the same mass and radial mass distribution as the pure NS, it is not surprising that the central pressure is the same as well.

Starting with an equilibrium NS and adding a small mPBH, to achieve an equilibrium solution with the same total mass, we find that the Hawking pressure at the corona must be equal to the original NS central pressure: $P(r_C) = P_0$.  For the BH to also be in equilibrium, this condition dictates a required accretion rate:
\begin{equation}
\label{eq:NS_equilibrium_condition}
   \dot E_\mathrm{acc}^\mathrm{eq} = c A_c P_0 
\end{equation} 
If black hole accretion rate does not match $\dot E_\mathrm{acc}^\mathrm{eq}$, the mPBH-NS will not be in hydrostatic equilibrium, with the central pressure too high or too low for the given amount of mass.

If the accretion rate does match \eqref{eq:NS_equilibrium_condition} and the mPBH-NS is in equilibrium, then $P(R_c) = P_0$ which determines the non-extremality parameter $x$ via \eqref{eq:equilibrium_condition}.  The mPBH will have an equilibrium temperature
\begin{equation}
\label{eq:BHTemp_equilibrium}
   T_\mathrm{BH} = \left(8.7 \times 10^3 \ \rm \frac{K}{Pa^{1/4}}\right) P_0^{1/4}
\end{equation} 
For the fiducial $1.35 M_\odot$ NS, this yields $T_\mathrm{BH} = 5.8\times 10^{12} \ \rm K$.

In solving the TOV equations \eqref{eq:ode1} and \eqref{eq:ode2} we have neglected temperature, but the NS heats up due to the Hawking radiation. Because the NS interior is a superfluid, heat is rapidly transmitted radially outward from the core. Once the NS reaches thermal equilibrium, the energy emitted at the corona into the NS must be emitted at the NS surface at an equal rate. From the Stefan-Boltzmann law the power radiated by thermal radiation is $P \propto A T^4$.  Equating the incoming Hawking power with the outgoing thermal radiation at the NS surface gives
the NS temperature: 
\begin{equation}
\label{eq:EquilibriumTemperature}
    T_\mathrm{NS} = 
    \sqrt{\frac{r_c}{r_\mathrm{NS}}}T_\mathrm{BH} = \left(1.1 \times 10^{-7} \ \rm \frac{K}{Pa^{1/4}}\right) Q^{1/4} P_0^{1/4}
\end{equation}
where, the equilibrium temperature of the BH is given by \eqref{eq:BHTemp_equilibrium}.  The fiducial $1.35 M_\odot$ NS reaches an equilibrium temperature of $T_\mathrm{NS} = (73\  \rm K) Q^{1/4}$ .

The approximation that the NS temperature has negligible contribution to its structure clearly breaks down when the NS temperature approaches the Fermi temperature $T_F$, the greatest temperature at which neutron star matter can remain degenerate. The Fermi temperature for the NS is given by
\begin{equation}
    k_B T_F = \epsilon_F = \frac{\hbar^2}{2m_n} 
    \left( 3\pi^2 n_n \right)^{2/3},
\end{equation}
where $\epsilon_F$ is the Fermi energy \cite{Carroll2007}, and $m_n$ and $n_n$ are the mass of a neutron and number density of neutrons, respectively. For the neutrons in a $1 M_\odot$ NS of radius 10~km, assuming a uniform density star, the Fermi temperature is $T_F\approx5\times10^{12}$~K. 

The requirement that the equilibrium NS does not get too hot implies a bound on the Hawking radiation rate, and therefore on the charge $Q$:
\begin{equation}
\label{eq:MaxCharge}
    Q \lesssim 10^{40}
\end{equation}

This equilibrium configuration in which the mPBH accretes matter from the NS which it then re-emits as Hawking radiation is in fact only quasi-stationary, with a finite lifetime.  Eventually the mPBH will accrete an order-one fraction of the NS mass.  From the accretion rate \eqref{eq:NS_equilibrium_condition}, the time to accrete a solar mass is approximately
\begin{equation}
\label{eq:lifetime}
\tau = 10^{31} {\rm \ yr} \ Q^{-2}.
\end{equation}

\subsection{mPBH-NS scenarios}
\label{sec:scenarios}

The final state resulting from the mPBH collision with a NS depends on the charge (or equivalently the extremal mass) of the black hole.  Three qualitatively distinct outcomes are possible, depending on the size of the magnetic charge. As we will see, only a NS that absorbs a black hole with a small magnetic charge can form a stable system. Above a certain threshold, the mPBH can not produce enough Hawking radiation to support the NS, and the entire NS will eventually collapse into the black hole.

For definiteness, we will consider a mPBH colliding with a our fiducial $1.35 M_\odot$ NS.

\subsubsection{Small Charge: $Q < 2.2 \times 10^{17}$}

We first consider the case of a mPBH with small charge, $Q < 2.2 \times 10^{17}$.  Just after the collision, once the mPBH has settled down to the center of the NS, it begins accreting NS matter. In the absence of numerical studies of the accretion by a mPBH inside a NS, we will assume the accretion rate is approximately the same as that of an uncharged PBH \eqref{eq:ELaccretion}. In terms of an energy accretion rate, this is
\begin{eqnarray}
\label{eq:ELaccretion_energy}
\dot E = 4.1 \times 10^{43} \ {\rm \frac{J}{m^2 s}} \  A_\mathrm{PBH} =  5.1 \times 10^{-25} \ {\rm \frac{J}{s}} \  Q^2
\end{eqnarray}
As the mPBH them becomes non-extremal, it will start to emit Hawking radiation, and the resulting outward radiation pressure would be expected to somewhat reduce the accretion rate. The uncharged accretion rate \eqref{eq:ELaccretion} or \eqref{eq:ELaccretion_energy}, which does not include Hawking radiation, then gives an upper bound on the initial accretion rate.  

As described above in \ref{sec:BH_equil}, the mPBH reaches equilibrium when the Hawking luminosity equals the accretion rate.  Using the uncharged accretion rate, the equilibrium condition \eqref{eq:equilibrium_bound} gives:
\begin{equation}
\label{eq:equiibrium_F}
   F(x)^4 =  \left(4.4 \times 10^{80}\ {\rm \frac{J}{m^2s}}\right) \ Q^{-3}  
\end{equation}
As long as $Q < 10^{20}$, a solution to \eqref{eq:equiibrium_F} exists, with $x \leq x_c$.

\begin{figure}
    \centering
    \includegraphics{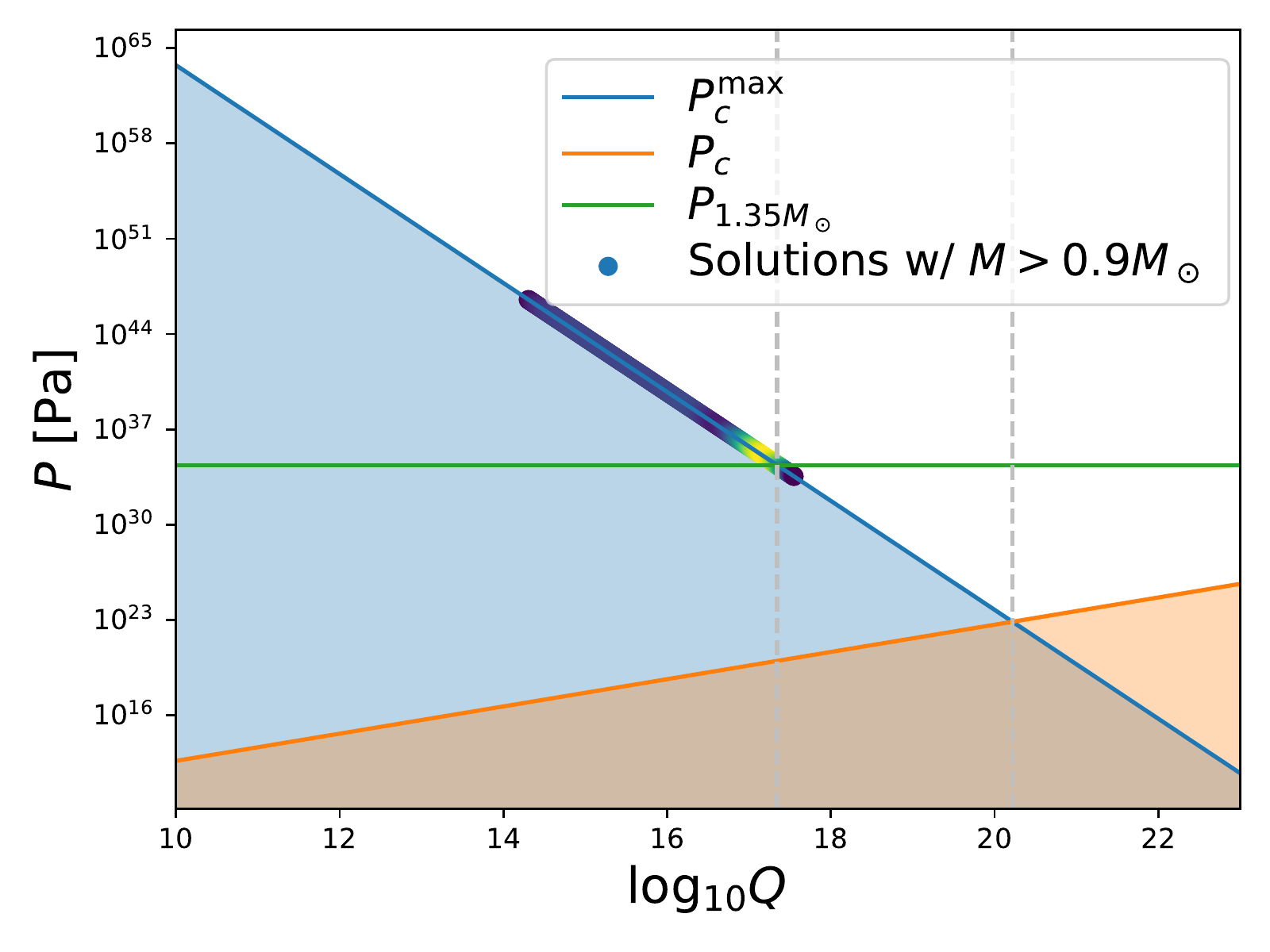}
    \caption{Configuration space of PBH+NS solutions. The blue line gives the maximum pressure $P_c^\mathrm{max}$ \eqref{eq:pressure_bound}. The orange line gives pressure $P_c$ resulting from the initial accretion rate \eqref{eq:EL_pressure}. Particular PBH+NS solutions with masses above $0.9M_\odot$ are shown with points whose color is determined by their mass. Solutions that support a fiducial mass of $1.35M_\odot$ have roughly constant pressure at the corona radius and are shown with a horizontal green line.}
    \label{fig:pressure}
\end{figure}

The pressure generated by the resulting Hawking radiation is then
\begin{equation}
\label{eq:EL_pressure}
    P_c =  \left(450 \ {\rm Pa}\right) Q 
\end{equation}
as shown in Fig. \ref{fig:pressure} as the upward-sloping orange line.  However, for a fiducial $1.35 M_\odot$ NS, the needed central pressure is $P_c = 2.37 \times 10^{34} \ \rm Pa$, which is shown in Fig. \ref{fig:pressure} as a horizontal green line.  Consequently, for $Q \leq 10^{33}$, the pressure being generated by this Hawking radiation is too low to yield an equilibrium NS solution.  Numerically simulating the non-equilibrium dynamics is beyond the scope of this paper.  However, it is reasonable to suppose that, without sufficient outward pressure, the NS core will start to collapse, and as this happens, the accretion rate will increase.

As the accretion rate increases, the resulting Hawking luminosity and resulting radiation pressure likewise increase.  In Fig. \ref{fig:pressure}, the system rises vertically from the red line. However, as shown in Fig. \ref{fig:Hawkingflux}, there is a maximum luminosity and corresponding maximum pressure, which occurs at $x_c = 0.15$.  From \eqref{eq:equilibrium_bound}, the maximum radiation pressure is given by
\begin{equation}
\label{eq:pressure_bound}
    P_c^\mathrm{max} =  \left(5.6 \times 10^{106} \ {\rm Pa}\right) Q^{-4} 
\end{equation}
This upper bound on the Hawking pressure is shown in Fig. \ref{fig:pressure} as the downward-sloping blue line.
For the small Q regime, $Q < 2.2 \times 10^{17}$, the accretion rate and Hawking luminosity can rise until the mPBH-NS equilibrium is reached.  

In lieu of a dynamical simulation of this process, we can estimate the time for the accretion rate to increase and raise the Hawking pressure enough to stabilize the NS.  An internal dynamical time scale for the NS  can be set by speed of sound, which is determined by the equation of state \eqref{eq:eos}. At high density the conformal symmetry of QCD is restored, and the speed of sound in the NS core can be approximated by the conformal value $v_s \sim c/\sqrt{3}$ \cite{Altiparmak:2022bke}.  The sound crossing time $\tau_s = R_\mathrm{NS}/v_s \sim 10^{-4} \, \rm s $ gives a lower bound on the time for the accretion rate to increase.  This process is slow compared to the microscopic timescale on which the mPBH reaches equilibrium, but short compared to any astrophysical timescales.

Once the mPBH-NS system has reached equilibrium, the lifetime of this quasi-stationary state is approximated by \eqref{eq:lifetime}, the time for an order one fraction of the NS to be accreted and re-radiated by the mPBH, which scales as $Q^{-2}$.  For $Q < 10^{10}$, the lifetime is longer than the age of the universe: $\tau > 10^{11} {\rm \ yr}$.  However, for $Q \lesssim 10^{17}$, near the upper end of this range, the lifetime is quite short $\tau \gtrsim 10^{4} {\rm \ s}$.

\subsubsection{Medium charge: $2.2 \times 10^{17} < Q < 1.7 \times 10^{20}$}
For a somewhat larger magnetic charge, the evolution of the mPBH-NS system begins similarly to the small $Q$ regime discussed above.  The mPBH  begins accreting at the initial rate \eqref{eq:ELaccretion} and quickly comes to equilibrium, generating an outgoing Hawking pressure $\eqref{eq:EL_pressure}$, which is insufficient to yield an equilibrium NS.  

Again, we expect the out-of-equilibrium NS to begin collapsing, driving up the accretion rate, which in turn causes the black hole mass and the Hawking pressure to increase.  However, the necessary central pressure to support the NS is not possible.  As can be seen in Fig. \ref{fig:pressure}, in this medium-$Q$ regime, the maximum Hawking pressure $P(r_c)^{max}$ (the blue line) is below the necessary pressure to support a $1.35 M_\odot$ NS (the green line).  Starting with the initial Hawking pressure $\eqref{eq:EL_pressure}$ (the orange line), the pressure increases until the maximum $P(r_c)^{max}$ is reached, which is still insufficient for the NS reach equilibrium.  As in the small-$Q$ case, the timescale for this dynamical process is driven by the NS speed of sound; we estimate the time scale for the accretion rate to increase to be $\tau \gtrsim 10^{-4} \, \rm s$.

The accretion rate continues to grow, but now, as the mPBH mass continues to grow, the temperature and the luminosity begin to fall, with the mPBH crossing over to Schwarzschild-like behavior.  At this point, the matter begins falling in to the mPBH faster than it can be remitted,  black hole is no longer in equilibrium, and the system becomes unstable.  The subsequent accretion and black hole growth closely resemble that of an uncharged black hole, as described in \cite{East:2019dxt, Richards:2021upu}, and the entire NS is eventually consumed by the black hole in a time $\tau \sim \left(5\times 10^{25} \, \rm yr\right) Q^{-1}$ \cite{Baumgarte:2021thx}, which for this range of $Q$ gives a lifetime between $\tau \sim 10^{5}\, \rm yr - 10^{8} \, \rm yr$.
The final stages of the NS collapse can be a catastrophically violent event, with the possibility of emitting large amounts of matter or radiation, depending on the details of the collapse process.

In the end, there remains a stellar-mass, magnetically charged black hole.  Unlike the original mPBH, this black hole is far from extremal, with a very small Hawking temperature, but the original magnetic charge remains.

\subsubsection{Large charge: $Q > 1.7 \times 10^{20}$}
For large charge $Q > 1.7 \times 10^{20}$, Fig. \ref{fig:pressure} would seem to indicate that the initial accretion rate \eqref{eq:ELaccretion} generates a Hawking pressure larger than $P_\mathrm{max}$ which, of course, is not possible.  Instead, at an accretion rate \eqref{eq:ELaccretion}, the NS matter is falling across the black hole horizon faster than the Hawking radiation can emit it the energy back out again.  As a result, the black hole never reaches equilibrium.  Starting with a highly charged extremal mPBH, the mass grows, increasing the Hawking radiation, until the maximum luminosity at $x_c = 0.15$ is passed and the black hole crosses over into Schwarzschild-like behavior.  

Beyond this point, the black hole mass continues to grow, now with the temperature and luminosity decreasing.  As a result, the accretion proceeds in much the same way as for an uncharged PBH, as  in \cite{East:2019dxt, Richards:2021upu}, with the black hole growing while accreting the entire NS in a timescale $\tau \sim \left(5\times 10^{25} \, \rm yr\right) Q^{-1} \lesssim 10^5\, \rm yr $ \cite{Baumgarte:2021thx}. As with the medium charge mPBH discussed above, the NS ultimately collapses into the BH, finally ending up as a magnetically charged solar-mass BH.

\section{Observables}
\label{sec:observables}

\subsection{Thermal radiation}

We now investigate the possibility of astronomically observing the results of a mPBH-NS collision.  In the previous section, we found that the collision of a NS and a mPBH with small charge will result in a long-lived, quasi-stable mPBH-NS bound state.  The observational challenge is to distinguish it from an ordinary NS without a mPBH. 

As discussed above, Hawking radiation from the mPBH heats up the NS which, as a result, emits thermal radiation out into space.  At thermal equilibrium, the NS temperature is given by \eqref{eq:EquilibriumTemperature}
\begin{equation}
    T_\mathrm{NS}  = 1.5 \times 10^6 \, \rm K \left(\frac{Q}{2.2\times 10^{17}}\right)^{1/4}.
\end{equation}

An ordinary NS, however, has a non-zero temperature and also emits thermal radiation. In particular, a recently-formed NS typically has a high temperature. For example, the many examples in \cite{Page:2004fy} have $T \gtrsim 10^6 \, \rm K$.  In this case, the NS is already so hot that the additional heat, even from a near maximal $Q \sim 10^{17}$ mBPH, would likely not be observable.

However, as a NS ages, it cools, and can cool  below 1000K in a billion-year time scale \cite{Yakovlev:2004iq, Page:2004fy}. The additional thermal radiation from a mPBH with $Q \gtrsim 10^4$ could then be observable. Assuming the age of the NS can be reliably estimated from, for example, the spin down rate, a higher than expected temperature could indicate the presence of a radiating mPBH inside.

Sufficiently old and nearby neutron stars can be detected by pulsar emission via radio telescope, such as FAST or the VLA.  Follow up observation by infrared telescope, such as the James Webb Space Telescope, the Thirty-meter Telescope, or the Extremely Large Telescope, could then be able to accurately measure the temperature \cite{Raj:2017wrv}. 

There are alternative mechanisms, however, by which a NS could be heated.  For example, dark matter accumulating and self-annihilating in the core can also generate significant heating, up to as much as $2500 \, \rm K$ \cite{Raj:2017wrv, Garani:2020wge}. Although the thermal signal from a $Q\lesssim 10^6$ mPBH might closely resemble that due to dark matter accretion, a mPBH with $Q > 10^8$ would be sufficiently high temperature as to exclude dark matter as the heat source.

\subsection{Magnetic field}
\label{sec:magnetic_field}

A particularly distinguishing feature of the mPBH is its characteristic magnetic field. Near the horizon, the magnetic field is very strong. At the corona, the magnetic field is at the value which the electroweak symmetry is restored, $B_c = B_\mathrm{EW}= 10^{20} \, \rm T$ \cite{Bai:2020spd}. Assuming the magnetic field of the mPBH outside the corona is a spherically-symmetric monopole field, then $B \propto Q/r^2$.  
Taking the radius of the NS to be $r_\mathrm{NS} = 10^4\,  \rm m$, the magnetic field at the surface is only
$$B_{\rm surface} \approx (10^{-24} \ {\rm T})\,   Q \lesssim 10^{-7}\ {\rm T} \ .$$
Compared with the large magnetic field of a NS, which is typically at least $10^4 \, \rm T$, the contribution due to the mPBH would appear completely negligible.

In the core of the NS, however, the magnetic field of the mPBH is quite strong.  Although, we have so far mostly neglected the effect of the mPBH's magnetic field on the NS, it could significantly affect the NS matter in the core and change the structure of the NS.  Strong magnetic fields could affect the equation of state, in particular the hadron-quark phase transition or transitions to hyperon or other exotic phases. 

For example, the interior of the NS is expected to feature proton superconductivity, implying the magnetic field of the mPBH outside the corona would be confined to one or more narrow flux tubes \cite{Lander_2013, Graber_2015}.  As a result, the magnetic field could be highly concentrated in a few spots, causing significant deformation of the NS magnetosphere from a dipole.  
Interestingly, the NICER telescope recently observed deviations from a simple dipole in the surface field of a NS~\cite{Bilous_2019,Miller:2019cac}.

However, the ability to make statements regarding observational consequences of the mPBH magnetic field relies on a solid understanding of an  NS without a mPBH.  Currently, even for an ordinary NS, neither the geometry nor the strength of the internal magnetic field are well understood.  The affects of the NS's own magnetic field on the equation of state and internal structure are currently under investigation \cite{Soldateschi:2021cxq}.

\subsection{Collapse}
\label{sec:collapse}

For a mPBH with a medium or large charge, $Q>10^{17}$, the NS eventually collapses into the mPBH, a highly-energetic event which would likely have observable consequences.  For example, the collapse of the NS magnetosphere into a BH could lead to a large transient flux of electromagnetic energy which has been identified as a potential FRB source \cite{Fuller_2015, Abramowicz:2017zbp}.  In particular as noted in Ref.~\cite{East:2019dxt}, simulations of the collapse of magnetized neutron stars demonstrate significant electromagnetic radiation~\cite{Lehner:2011aa,Falcke:2013xpa}. Generally a dipole field is assumed, whereas here there would be a significant monopole component which likely would change the geometry of this emission. That the monopole field would not be shed during collapse would also be significant, as discussed in the next section. Other possible signatures include kilonavae \cite{Bramante:2017ulk}, gamma-ray bursts \cite{Chirenti:2019sxw}, and gravitational waves \cite{Baiotti:2007np, Kurita:2015vga, Abramowicz:2017zbp}.

However, the collision of an uncharged PBH with NS also leads to a the NS collapsing into the BH.  Potentially, a NS collapse into a mPBH can be observationally distinguished, but it would require a more detailed understanding of the late collapse process and the mechanism by which the FRB is generated.  When a mPBH collides with a NS the interaction of their magnetospheres could lead to an FRB-like prompt radio pulse. This sort of emission mechanism has been investigated in the case of a NS-NS collision \cite{Pshirkov_2010}. But the mPBH-NS collision would be different, in detail. Of particular importance would be the strength of the mPBH magnetic field, dependent on $Q$. Further investigation of this scenario is warranted.

\subsection{Magnetic black hole}
\label{sec:magneticBH}

The final end product of this collapse is a magnetically charged, solar-mass black hole.  Compared with black holes formed by stellar collapse, this magnetic black hole would be rather light, with a mass $M \lesssim 2 M_\odot$ smaller than any observed astrophysical black hole. But, compared to the original, microscopic mPBH, such an object could be observed in several ways.  A rapidly spinning black hole in a magnetic field can convert rotational energy into jets of electromagnetic radiation via the Blandford-Znajek process \cite{Blandford:1977}.  Typically, the magnetic field is generated by accreting matter, but, in this case, the magnetic field is sourced by the black hole itself.  

This has a couple potential implications.
First, the black hole could produce Blandford-Znajek jets in the absence of any accretion disk.
Alternatively, if the magnetically charged black hole were in a binary with another black hole, the orbital motion could also result in electromagnetic jets~\cite{Liebling:2016orx}.\footnote{The distinct gravitational wave signal from a such a binary was investigated in \cite{Takhistov:2017bpt}.} This contrasts
with studies of uncharged black hole mergers within an externally sourced magnetic field~\cite{Palenzuela:2010nf}. In particular, black holes dynamically shed any magnetic field higher than the monopole in accordance with the no hair conjecture.

\section{Discussion}
\label{sec:discussion}

Compared with the collision between a NS and an uncharged BH, the collision of a NS with a magnetically charged BH can have very different dynamics and have an entirely different outcome.  
Due to electromagnetic interactions, a charged BH stops almost immediately and transfers its angular momentum to the NS, resulting in a glitch or anti-glitch. If the NS is observed as a pulsar, current instruments are only sensitive enough to detect the collision of a mPBH with very large charge. Future gravitational wave detectors, however, might be able to detect a  wider range of collisions.

After the collision, different thermodynamics of charged BHs and the region of restored electroweak symmetry just outside the horizon means that the NS does not inevitably collapse into the BH. For small enough magnetic charge, we argued the mPBH and NS can reach a long-lived quasi-equilibrium state in which accretion of NS matter into the BH generates Hawking radiation which then supports the NS.  Black holes with large magnetic charge, perhaps surprisingly, are not able to generate enough pressure to support the NS, in which case the entire NS is eventually accreted by the BH, much like an uncharged BH.  Observationally, it may be challenging, but not impossible to distinguish a pure NS from one with a small-charge mPBH at its core.  Similarly, a NS collapse due to a large-charge mPBH may look quite similar to that due to an uncharged PBH, and further investigation is required to determine the differences.  Furthermore, the resulting magnetically charged solar-mass BH could be more readily observed. 

We should note that our treatment of the mPBH-NS system is highly simplified. First, we largely ignore the detailed interplay of the Hawking radiation with the neutron star matter, which is very complicated.  In particular, we assume that the Hawking radiation is readily absorbed by the NS matter, but we neglect the significant radiation pressure on the infalling matter.  We expect this would decrease, perhaps significantly, the accretion rate. 

Second, the dynamics of the stellar material near the black hole is extremely complicated involving a huge magnetic field, the intense spectrum of particles radiated via the Hawking process, and high temperatures (and concomitant nuclear reactions). Instead, as a first step, we generally ignore these effects, expecting them to be subdominant to fluid pressure and gravity, at least for the qualitative aspects of the equilibrium solutions. For example, it is known that a very large magnetic field alters the equation of state of dense matter in such a way as to support larger masses than with no magnetic field \cite{Watanabe:2022cuv,Broderick:2000pe,Suvorov:2021ymy}.   As mentioned in Sec. \ref{sec:magnetic_field}, the large magnetic field may also form flux tubes which could significantly deform the neutron star away from spherical symmetry. 

Third, we need to better understand the out-of-equilibrium dynamics of the NS, specifically in the case where the Hawking radiation is insufficient to support the NS.  We assumed, in this case, that the accretion rate would increase.  However, a more sophisticated treatment, for example, a dynamical numerical simulation, would probably be needed to study the NS evolution.

In modeling mPBH, we assumed most of the Hawking radiation emitted from the horizon was stopped at the corona, where the electroweak symmetry breaks and the fermions become massive.  The corona was modeled as a hot radiating sphere in thermal equilibrium with the black hole.  However, an improved picture of the Hawking radiation is needed. For example, some light particles may pass directly through the corona.

A further limitation is that we treated the NS matter in a fluid approximation, which breaks down around the neutron length scale of $10^{-15} \rm m$, while we treated the mPBH as a macroscopic object in that fluid.  However, the mPBH can be extremely small, particularly for small charge; the corona radius, for example, is $r_c = 1.7 \times 10^{-18} {\rm \ m} \ \sqrt{Q}$.  For $Q < 10^6$, this is smaller than the radius of a neutron and our results would likely be untrustworthy.  Instead, the mPBH would need to be considered as a microscopic object interacting with individual nucleons, as discussed in \cite{Giddings:2008gr}.

\begin{acknowledgments}
This work is supported by the National Science Foundation via grants PHY-1912769 and PHY-2011383~(SLL), PHY-2000398~(ML), AST-2011757~(JHS), and AST-2011731~(MK).
Computations were performed at XSEDE.
\end{acknowledgments}

\bibliographystyle{utphys}
\bibliography{paper}

\end{document}